\documentclass[12pt]{article}
\usepackage{a4}
\usepackage{epsfig,graphicx}

\newcommand\ignore[1]{}

\begin{document}

\begin{center}
{\Large \textbf{Topological Character of Excitations in }}

{\Large \textbf{Strongly Correlated Electronic Systems: }}

{\Large \textbf{Confinement and Dimensional Crossover. }}

\bigskip 

\textbf{S. Brazovskii }

\bigskip 

\end{center}

\textit{Laboratoire de Physique Th\'eorique et des Mod\`eles Statistiques,
CNRS,}

\textit{B\^{a}t.100, Universit\'{e} Paris-Sud, 91405 Orsay cedex, France.}
\textbf{\footnote{
Also\textit{\ L.D. Landau Institute, Moscow, Russia. }}}

\textit{brazov@ipno.in2p3.fr\qquad\
http://ipnweb.in2p3.fr/~lptms/membres/brazov/ \bigskip }

\bigskip 

\begin{center}
Abstract
\end{center}

\bigskip 

Topologically nontrivial states are common in symmetry broken phases at
macroscopic scales. Low dimensional systems bring them to a microscopic
level where solitons emerge as single particles. The earliest and latest
applications are conducting polymers and spin-Peierls chains. After a
history introduction, we shall discuss the topological aspects of elementary
excitations, especially the confinement and the dimensional $D$ crossover.
At the $1D$ level we exploit results of different exact and approximate
techniques in theory of solitons, for both quantum and semiclassical models,
and the related knowledge on anomalous charges and currents. At higher $D$
the topological requirements for the combined symmetry originate the spin-
or charge- roton like excitations with charge- or spin- kinks localized in
the core. In quasi $1D$ world they can be viewed as resulting from a
spin-charge recombination due to the $3D$ confinement. 

\bigskip 

\section{History Introduction.}

This publication is not a regular article but rather a script of the recent
talk at \cite{20years}, p. 169, with extensions from the one at \cite
{Poincare}. Thus there will be no systematic references except for reviews,
but I shall try to remind the names. Self-references on related publications
can be found at the author's web site. The historical part
\footnote{
A specific for the anniversary discussion of \cite{20years} is shortened in
the present version.} 
is guided by real time observations since three
decades which differ from contemporary views. The original part is largely
unpublished: it was developed through a sequence of talks starting from the
lecture at \cite{nobel}. First will be an excursion to a relevant history of
interacting fermions in quasi one-dimensional -$1D$ systems. Topics on
solitons and their origin will follow: classical, quasi-classical and
quantum models, relations to field theory; 1D features - anomalous charge,
spin, currents, nontrivial topology. Finally we shall come to the goals -
problems at higher dimensions $D>1$: confinement, topological constraints,
combined symmetry, spin-charge recombination.

The discovery of Organic Superconductivity by Bechgaard, Jerome and Ribault
was indeed a long awaited D-day of landing. Among many earlier efforts in
chemistry and experiments, this anniversary recalls also the history of
theories for interacting electrons in (quasi) $1D$ systems. The roots hide
deep in decades winding through works of Bethe, Bloch, Fr\"{o}hlich,
Peierls, Tomonaga. The stem emerged some three decades ago, largely
nourished by the Little idea of a High-$T_{c}$ superconductivity in
hypothetical chain compounds. The three major lines were opened at once:
Bethe Anzatz exact solutions -- Gaudin, Lieb \& Wu; Bosonization --
Luttinger, Mattis \& Lieb, Luther \& Emery; Parquet Equations (an old form
for the later renormalization group RG) -- from Bychkov, Gor'kov \&
Dzyaloshinskii till Dzyaloshinskii \& Larkin and Solyom \textit{et al}. By
late 70's the prophets have left the field leaving for us the Old Testament (
\cite{Solyom}, \cite{Emery}) which later became obsolete and by now
abandoned in favor of New ones. Nowadays only a semantics like the
\textit{g-ology }reminds this breakthrough epoch.

Physics of microscopic solitons is a late branch of theoretical studies in
1D models. It was endorsed by high expectations from non perturbative
methods and by general interests in topologically nontrivial configurations,
e.g. \cite{solitons,raja}, see more refs. in \cite{Kirova}. First were
classical models with degenerate ground states (Bishop, Krumhansl, Rice,
Schrieffer, \textit{et al}). Largely the studies were stimulated by first
observations of Charge Density Waves -CDW's (Comes, Pouget, Shirane 
\textit{et al}) and especially by discovery of their sliding (Monceau \& Ong) where
effects of commensurate versus incommensurate (C, IC) states seemed to be
important{\footnote{
As a solitonic lattice the slightly IC state has been first described by
Dzyaloshinskii for magnetic media. Later it was rediscovered several times,
e.g. for CDWs.}. An impact came also from the field theory where quasi
classical solitonic solutions of multi-electronic Gross-Neveu type models
were discovered (Neveu \textit{et al}, Fateev). The multi-electronic theory
of adiabatic models (Peierls, Fr\"{o}hlich) in applications to CDW/SDW was
developed \cite{SB-84,SB-89,SB-96}. A very popular discrete version of the
Peierls model is known as the Su-Schrieffer-Heeger one \cite{polymers,Yu-Lu}
, in which frame much numerical work on solitons have been performed. A
surprise from these models with seemingly well behaved normal electronic
spectra in the ground state was that their excited states developed
spin-charge and spectral anomalies \cite{SB-84}. It was a consequence of
electronic selftrapping to topologically nontrivial forms: amplitude and
mixed solitons. The discovery of doped polymers has triggered an avalanche
of experimental (especially initiated by Heeger \textit{et al}) and related
theoretical (Schrieffer \textit{et al}, S.B.-Kirova-Matveenko; Bishop \&
Campbell, Maki \textit{et al}, Yu Lu \textit{et al}) studies on solitons,
see reviews \cite{SB-84,polymers,Yu-Lu}. The key words were solitons,
polarons, bipolarons, mid-gap states, pseudo-gap, etc. }

The contribution from our place was initially provoked by the problem, posed
that time by Lee, Rice and Anderson, of a pseudo-gap 
\footnote{
An essential part of these results (S.B. \& Dzyaloshinskii 76, S.B. 78,80)
have been recently republished, in a striking similarity, by the
Ohio-Australia group but unfortunately only one side of the old studies have
been captured. The missing link is a division of characteristic time scales:
fast like in Optics and PES against slow like in kinetics and
thermodynamics. Only the fast ones are relevant for electronic pseudo gaps
while the slow ones must be signatures of either collective modes or of
associated solitons \cite{SB-96}.}
Largely the studies were based on finding exact solutions for major
basic models and later on realizing the confinement problem. We benefited
from the field theory results (Neveu \textit{et al}, Fateev) and from
special mathematical methods (Dubrovin \& Novikov, Krichever). On this way
the solitonic solutions were generalized to multiple quasi periodic
structures allowing to describe a simultaneous effect of the doping and the
spin (or interchain) splitting\footnote{
This theory is a route to current studies of spin-Peierls systems in high
magnetic fields (Berthier et al, Boucher \& Regnault in \cite{schegol} p.
1939, Poilblanc et al), especially if extended to doped case.} or the effect
of lattice discreteness, a spill-over formation (S.B., Gor'kov and
Schrieffer) of superlattices due to electronic 3D dispersion. Finally the
quantization of solitons was performed in these multi fermionic models (S.B.
\& Matveenko 89) showing new anomalous currents.

A fully quantum approach to solitons have been called earlier when an
equivalence had been found between models of interacting electrons,
Sine-Gordon and Coulomb gas, see e.g. \cite{tsvelik}. The most powerful tool
became the Bosonization, see e.g. the recent book \cite{gogolin}, especially
after the Luther \& Emery work. It allowed to access explicitly the
spin-charge separation, to refine the form of the Green functions -
logarithmic corrections upon the Dzyaloshinskii-Larkin scaling solution
(Finkelstein), etc. Both the RG and exact Bethe Anzatz solutions (Nersessyan
\& Wiegman) of the quantum Sine Gordon model have been employed, providing
the quantum description of the C-IC transition.

With solitons being inhabitants of the $1D$ underground, the questions were
risen (S.B. 1980): Can they cross the boarder to the higher $D$ world? If
yes, then are they allowed to bring their anomalies? A password was found -
a \emph{\ confinement}: as topological objects (see \cite{Kirova} for
references) connecting the degenerate vacuums, the solitons at $D>1$ acquire
an infinite energy unless they reduce or compensate their topological
charges.

Here we can notice a similarity in histories of solitons and of interacting
electrons with respect to entering the higher dimensional world. In both
cases a long staying and still practicing belief is that the exotic $1D$
properties will be maintained which is far from being correct in general.
For solitons this is a devastating effect of the confinement (S.B. 80). For
electrons this is a decoherence of their interchain hopings which erases the
memory of $1D$ correlations in the gapless (the Luttinger Liquid of later
days) regime (S.B. and Yakovenko 85).

\section{Confinement and combined excitations.}

\subsection{A quasi classical commensurate CDW: \newline
confinement of phase solitons and of kinks.}

Being a case of spontaneous symmetry breaking, the CDW order parameter 
$\sim\Delta \cos [\vec{Q}\vec{r}+\varphi ]$ 
possesses a manifold of degenerate
ground states. For an IC CDW the global symmetry allows for arbitrary shifts
in phase $\varphi $ which originates the phenomenon of sliding discovered by
Monceau and Ong (recall theories by Fr\"{o}hlich; Bardeen \textit{et al};
Lee, Rice \& Anderson; S.B. \& Dzyaloshinskii). For an $M-$ fold
commensurate CDW the energy $\sim \cos [M\varphi ]$ reduces the allowed
positions to multiples of $2\pi /M$, $M>1$. The trajectories connecting them 
$\varphi \rightarrow \varphi +2\pi /M$ are the phase solitons -PSs, or 
''$\varphi $- particles'' after Bishop \textit{et al}. For a special case (e.g.
in the conjugated polymer $trans-CH_{x}$) of $M=2$, the $\varphi $ shift by $
\pi $ is replaced by an equivalent sign change of the amplitude, the kink $
\Delta \rightarrow -\Delta $. It matches the existence of amplitude solitons
-ASs even for the IC CDW without the topological selection (S.B. 78). If the
system is $1D$, or at least it is above the $3D$ or $2D$ transition
temperature $T_{c}$, then the local symmetry coincides with the global one.
In these regimes the symmetry is not broken and the solitons are allowed to
exist as elementary particles. But in the symmetry broken phase at $T<T_{c}$
, any local deformation must return the system to the same, or equivalent
(modulo $2\pi $), state. Otherwise the interchain interaction energy (with
linear density $F\sim \left\langle \Delta _{0}\Delta _{n}\cos [\varphi
_{0}-\varphi _{n}]\right\rangle $) is lost when the effective phase $\varphi
_{0}+\pi sign(\Delta _{0})$ at the soliton bearing chain $n=0$ acquires a
finite (and $\neq 2\pi )$ increment with respect to the neighboring chain
values $\varphi _{n}$. The $1D$ allowed solitons do not satisfy this
condition which originates a constant \textit{confinement force} $F$ between
them, hence the infinitely growing confinement energy $F|x|$. E.g. for $M=2$
the ASs should be \emph{bound to pairs of kinks}; for an arbitrary $M$ the
PSs should \emph{aggregate in trains} of $M$ solitons, thus accumulating an
allowed increment $2\pi $ of the phase. For an IC CDW at $D>1$, the local $
2\pi $ solitons appear as the only allowed increments of one chain with
respect to the surrounding ones. If a large number of adjacent chains
acquires this increment coherently, then one can recognize the dislocation
loop -DL embracing this bundle (Dumas, Feinberg, Friedel; S.B. and
Matveenko). The aggregation of solitons into minimal allowed complexes is
expected for high $T$ versus low concentration, for impurities trapping or
due to long range Coulomb interactions. For a pure system and at low $T$,
when the entropy plays no more role, the solitonic complexes should
aggregate into macroscopic objects, the bubbles. They are encircled by the
lines which are relevant to DLs in IC systems. But instead of the
homogeneous $2\pi $ circulation of the phase around the DL, the bubble
boundary emits the $M$ surfaces (stripes) each providing the $2\pi /M$ phase
jumps, just as it has been observed by Chen \textit{et al}. When the bubbles
extends to the sample width, the plains decouple into unbound Domain Walls
separating domains with different global values of the order parameter. It
was found (S.B. \& T.Bohr) that the crossover between the aggregation due to
the confinement and the size effect applies to $2D$ systems only. In $3D$ a
second phase transition of an infinite aggregation should be expected.

In summary, here are the known scenarios for the deconfinement:

\begin{description}
\item  Aggregation versus thermal deconfinement (S.B. \& T.Bohr 83);

\item  Coupling of kinks to structural topological defects - twistons in
polymers

(S.B. \& Kirova 88,91,92);

\item  Binding into kink-antikink pairs and the origin of bipolarons

(S.B. \& Kirova 81);

\item  A compensation by gapless degrees of freedom: the \emph{today's
specialty.}
\end{description}

\subsection{ A quasi classical CDW: confinement of Amplitude Solitons with
phase wings.}

The CDW order Parameter is $O_{cdw}\sim \Delta (x)\cos [Qx+\varphi ]$ where $
\Delta $ is the amplitude and $\varphi $ is the phase. In $1D$ the ground
state with an odd number of particles contains the AS $\Delta (x=-\infty
)\leftrightarrow -\Delta (x=\infty )$, Figure 1. It carries the singly
occupied mid-gap state, thus having a spin $1/2$ but its charge is
compensated to $0$ (fractional variable charges also appear at
circumstances) by local dilatation of singlet vacuum states. As a nontrivial
topological object ($O_{cdw}$ does not map onto itself) the pure AS is
prohibited in $D>1$ environment. Nevertheless it becomes allowed in $D>1$ if
it acquires phase tails with the total increment $=\pi $, Figure 2. The
length of these tails $\xi _{\varphi }$ is determined by a weak interchain
coupling, thus $\xi _{\varphi }\gg \xi _{0}$. As in $1D$, the sign of $
\Delta $ changes within the scale $\xi _{0}$ but further on, at the scale $
\xi _{\varphi }$, the sign of the factor $\cos [Qx+\varphi ]$ also changes
the sign thus leaving the product in $O_{cdw}$ to be invariant. As a result
the 3D allowed particle is formed with the AS core $\xi _{0}\sim 10\AA $
carrying the spin and the phase $\pi /2$ twisting wings $\xi _{\varphi }\sim
100\AA $, each carrying the charge $e/2$. Nevertheless the spin-charge
reconfinement is not the end of the story. When the combined soliton moves,
its locale electric current is compensated exactly by the back flow currents
at distant chains. This is the property of the vortex like configuration.
Finally, the a soliton as a state of the coherent media will not carry a
current and himself will not be driven be a homogeneous electric field. But
locally, e.g. interacting with a charge impurity, it will behave as a
charged particle. This paradox can be proved as a theorem (S.B. 80) for a
general system with electronic interactions. We shall return to this
property below.

\begin{figure}[h]
\begin{minipage}[c]{.45\linewidth}
\includegraphics*[width=5.5cm]{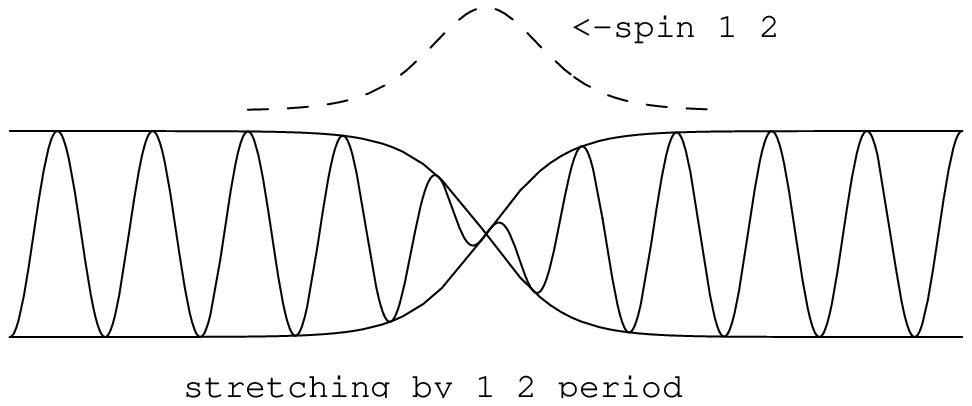}
\epsfxsize .7\hsize
\caption{An amplitude soliton in the IC CDW}
\end{minipage}\hfill 
\begin{minipage}[c]{.45\linewidth}
\includegraphics*[width=5.5cm]{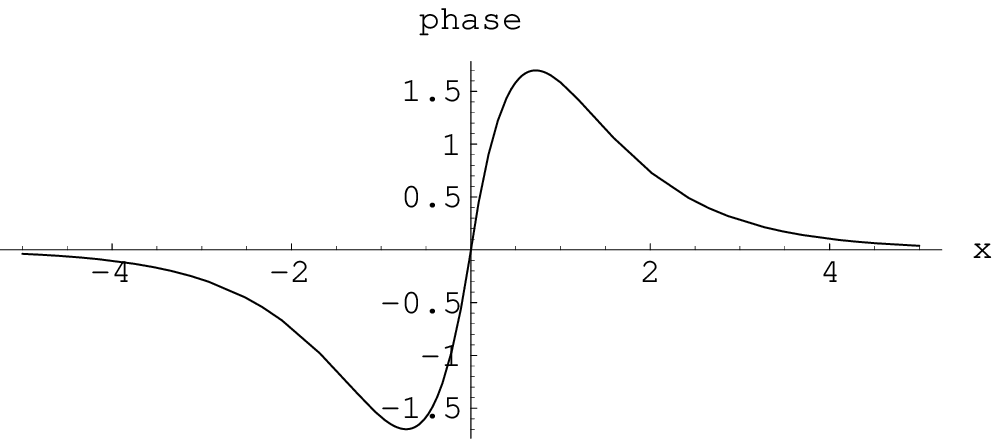}
\epsfxsize .35\hsize
\caption{Phase tails (between $0$ and $\pi/2$) adapting the AS.}
\end{minipage}\hfill
\end{figure}

\subsection{ Quantum models based on bosonization.}

\subsubsection{A hole in the AFM environment.}

Consider first the quasi $1D$ system with repulsion at a nearly half filled
band which is the Spin Density Wave -SDW rout to a general doped
antiferromagnetic -AFM Mott-Hubbard insulator. The bosonized Hamiltonian
(Luther \& Emery) can be written schematically (\emph{numerical coefficients
are not shown}, only terms relevant to repulsion are left) in $1D$ as 
\[
H_{rep}\sim \{(\partial \varphi )^{2}-Ucos(2\varphi )\}+(\partial \theta
)^{2}
\]
where $\varphi $ is the analog of the CDW displacement phase and $\theta $
is the angle of the $SU2$ spin rotation. Here $U$ is the Umklapp scattering
amplitude of Luther \& Emery (or $g_{3}$ of Dzyaloshinskii \& Larkin) which
is a feature of systems with a single mean occupation of lattice sites.
Normally $U$ is of the order of other interactions which are not small.
Hence a big gap is opened in the charge $\varphi -$ channel so that only
gapless spin degrees of freedom are left for observations. But a generous
feature of just the Bechgaard salts\footnote{
This feature was noticed in the very first work (S. Barisi\'{c} and S.B. 81)
done in response to the D-day which was a natural application of the
recently having been developed (S.B. and S. Gordyunin, 80) concept of the
''weak two-fold commensurability''. Even more intriguing manifestations of
various types of anions' ordering have been noticed and analyzed (S.B. and
V.Yakovenko 85) after systematic structural studies became available (see
J.-P. Pouget review talk in \cite{20years} and, with S. Ravy, in \cite
{schegol}).} is that $U\sim g_{3}$ is small appearing only as a secondary
effect of the anionic sublattice. It opens an intriguing crossover of the
charge localization - one of main subjects of current experiments (see in 
\cite{20years} the review talks by Jerome and Bourbonnais).

The degeneracy between $\varphi =0$ and $\varphi =\pi $ corresponds to the
displacement of the electronic system by just one lattice position $n$.
(Indeed, the wave functions at the Fermi points $\pm k_{F}$ are $\Psi _{\pm
,\sigma }\sim \exp [\pm i(n\pi /2+\varphi /2)]$, $\sigma =\uparrow
,\downarrow $). Hence the $\pm \pi $ soliton just adds/removes one electron
in real space. The excitations in $1D$~are the \textit{(anti)holon} as a $
\pm \pi $ soliton in $\varphi $ and the spin sound in $\theta $ which are
decoupled. At $D>1$ below the SDW ordering transition $T_{sdw}$ the order
parameter is 
\[
O_{sdw}\sim \Psi _{+\uparrow }^{+}\Psi _{-\downarrow }+\Psi _{+\downarrow
}^{+}\Psi _{-\uparrow }\sim \cos \varphi \,\exp \,\{i(Qx+\theta )\} 
\]
Following the reasons of the CDW case, we see that to survive in $D>1$ the $
\pi -$ soliton in $\varphi $, $\cos \varphi \rightarrow $ - $\cos \varphi $,
should enforce a $\pi $ rotation in $\theta $, then the sign changes of the
two factors composing $O_{sdw}$ cancel each other.

\begin{figure}[h]
\begin{minipage}[c]{.4\linewidth}
\includegraphics*[width=3.5cm]{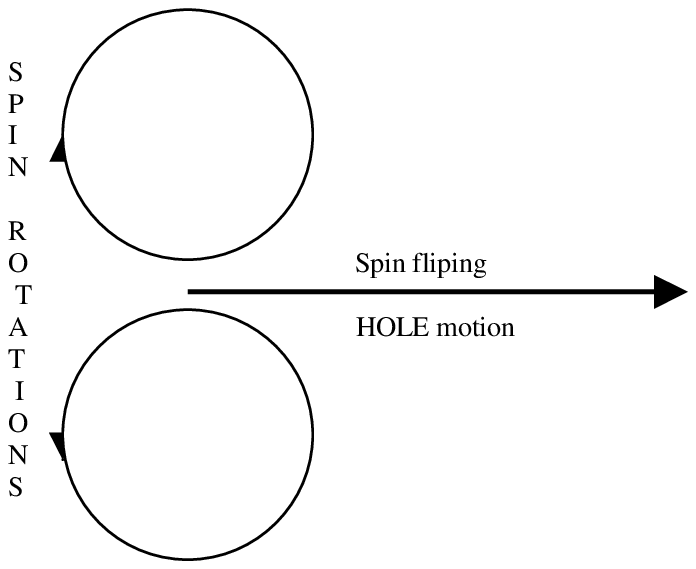}
\epsfxsize .85\hsize
\end{minipage}\hfill 
\begin{minipage}[c]{.6\linewidth}
{ Figure 3:} 
An illustration of the hole motion in the AFM media accompanied by the rotation of spins. 
The string of the reversed  AFM order created by the holon is cured 
by the semi-vortex of the staggered magnetization. 
The same figure is applied to the below case of the superconductivity 
with lines of electric currents instead of  contours of spin rotations and 
with the spin carrying normal core instead of the moving hole.
\end{minipage}\hfill
\end{figure}
\hfill

Now it is straightforward to build a universal formulation which extends to
isotropic AFM systems like the $CuO_{2}$ planes. The SDW is now a staggered
magnetization, the soliton becomes a hole moving within the AFM environment,
the $\pi $ wings become the magnetic semi-vortices (see \cite{Kirova} for a
macroscopic illustration). The resulting configuration is a half-integer
vortex ring of the staggered magnetization (a semi roton) with the holon
confined in its center, Figure 3. (In $2D$ the vortex ring is reduced to the
pair of vortices.) The solitonic nature of the holon in quasi $1D$
corresponds to the string of reversed staggered magnetization (Nagaev 
\textit{et al}, Brinkman and Rice) left by the moving hole (see \cite{brenig}
for the latest review). This approach allows to calculate the physical
variables like the localized charge, the delocalized staggered spin density $
S_{afm}$ and the net (ferromagnetic) magnetization $S_{\,fer}$, the
corresponding currents, etc. E.g. at large distances we have\footnote{
There is a typing mistake in the corresponding formula in \cite{20years}.} $
S_{afm}\sim {x/}({x^{2}+y^{2}})\;$where $x$ is the motion direction. The
induced non oscillating magnetization is 
\[
~S_{fer}\approx \delta (x)\delta (y)+\frac{1}{2\pi }{(x^{2}-y^{2})}/{
(x^{2}+y^{2})^{2}}
\]
Here the first term comes from the single spin density localized in the
wings of the combined core around the charged nucleus. The second term is
the counter rotation of the staggered magnetization at distant chains. The
coefficients of these terms are fixed and correlated in such a way that,
being integrated over the cross-section $y$, the two contributions cancel
each other \emph{identically in} $x$.\footnote{
To deal correctly with shape dependent dipole distribution, the prescription
is to interpret the last term as $\sim \partial _{x}[x/(x^{2}+y^{2})]$ which
invokes the semiroton as the origin of long range distortions.} This
cancellation, in addition to the combined core properties, is a new
information in addition to the spin wave theory results of Shraiman and
Siggia. It opens an important paradox already mentioned above for CDW:
integrally the spin current of the moving particle is transferred to the
collective mode of a homogeneous rotation leaving only the charge current as
for the holon in $1D$. But locally, via NMR e.g., it will be seen as the
nominal electronic spin. Thus in quasi-$1D$ picture, following the path of
the soliton bearing chain, the normal local quantum numbers are found: spin $
s=1/2$ (while in wings) and charge $e=1$ (while in the core).

\subsubsection{Spin-Gap cases: the quantum CDW and the superconductivity.}

For systems with attraction between electrons the bosonized Hamiltonian
reads 
\[
H_{atr}\sim \{(\partial \theta )^{2}-Vcos(2\theta )\}+(\partial \varphi
)^{2} 
\]
where $V$ is the backward exchange scattering (Luther \& Emery, $g_{1}$ of
Dzyaloshinskii \textit{et al}, see Gor'kov in \cite{schegol}). The Umklapp
term of the earlier $H_{rep}$ is either absent for an IC case or
renormalized to zero, in a similar way as it was for $V$ in the repulsion
case. The elementary excitations at $1D$ are the \textit{spinon} as a
soliton $\theta \rightarrow \theta +\pi $ and the gapless charge sound in $
\varphi $.

The CDW order parameter is 
\[
O_{cdw}\sim \Psi _{+\uparrow }^{+}\Psi _{-\uparrow }+\Psi _{+\downarrow
}^{+}\Psi _{-\downarrow }\sim \exp [i(Qx+\varphi )]\cos \theta 
\]
Hence for the CDW ordered phase in a quasi $1D$ system the allowed
configuration is composed with the spin soliton: $\theta \rightarrow \theta
+\pi $ and the wings $\varphi \rightarrow \varphi +\pi $ where the charge $
e=1$ is concentrated. Beyond a low dimensionality, a general view is: the 
\emph{spinon as a soliton bound to the semi-dislocation loop}. The same
paradox is observed as for the classical CDW and similar to the AFM cases:
locally the single electronic quantum numbers are reconfined (while with
different scales of localization); but integrally still there is only the
moving spin while the charge is transferred to the collective sliding mode.

For the singlet superconductivity SC the order parameter is 
\[
O_{sc}\sim \Psi _{+\uparrow }\Psi _{-\downarrow }+\Psi _{+\downarrow }\Psi
_{-\uparrow }\sim \exp [i\tilde{\varphi}]\cos \theta 
\]
where the gauge phase $\tilde{\varphi}$ is conjugated to the chiral one $
\varphi $ - the last term in $H_{atr}$ can be written reciprocally as $\sim
(\partial \tilde{\varphi})^{2}$ as well (Efetov \& Larkin). In $D>1$ the
elementary excitation is composed by the \emph{spin soliton} $\theta
\rightarrow \theta +\pi $ supplied with \emph{current wings} $\tilde{\varphi}
\rightarrow \tilde{\varphi}+\pi $. A quasi $1D$ interpretation is that the 
\emph{spinon works as a Josephson }$\pi -$\emph{\ junction} in the
superconducting wire. The $\ 2D$ view is a pair of superconducting $\pi -$
vortices sharing the common core where the unpaired spin is localized. The $
3D\ $ view is a half flux vortex loop which collapse is prevented by the
spin confined in its center.

In summary of this chapter notice that the factorization of typical order
parameters reflects the combined global symmetry of the ordered phase, e.g.
for the AFM, SDW this is a product of translations and the time reversal
(S.B. and I. Luk'yanchuk 89-91). Beyond applications to solitons, it affects
other observable physical properties dependent on spin-charge
reconciliation. Thus, the optical absorption across the spin gap becomes
allowed contrary to the basic models (S.B. and A. Finkelstein 81).

\section{Conclusions.}

The discovery of the organic superconductivity with its striking proximity
to the AFM phase has changed the accents of theories from the earlier
dominated Little model of strong on-chain attraction to the ones with
repulsion and the related Mott-Hubbard insulator phase. Adding the later
studies on the interchain electronic coupling/coherence we see that the
major problematics of the High-$T_{c}$ epoch has been risen years in
advance. Later on the theory of 1D electronic systems has been called to
elucidate the hypotheses risen in High-Tc epoch (P.W. Anderson). Here we
have explored this opportunity once again in some other aspects. Thus the
slightly IC SDW can be viewed as a prototype of the doped AFM, etc.
Exploiting another analogy, to quarks and gluons in QCD, separate spinons
and holons are not observed directly because of the topological confinement.
The topological effects of $D>1$ ordering reconfines the charge and the spin 
\emph{locally} while still with essentially different distributions.
Nevertheless \emph{integrally} one of the two is screened again, being
transferred to the collective mode, so that in kinetics the massive
particles carry only either charge or spin again.

\bigskip

\textbf{Acknowledgments.} I recall a great fortune of collaborations on
above subjects with S. Barisic, T. Bohr, I. Dzyaloshinskii, A. Finkelstein,
L. Gor'kov, N. Kirova, I. Krichever, A. Lebed, S. Matveenko, Ph. Nozieres,
J.R. Schrieffer and V. Yakovenko.

\end{document}